# Single-atom-at-a-time adsorption studies of $^{211}$Bi and its precursor $^{211}$Pb on SiO$_2$ surfaces

Dominik Dietzel[1,2], Alexander Yakushev[2], Christoph E. Düllmann[1,2,3], Jadambaa Khuyagbaatar[2], Jörg Krier[2], Egon Jäger[2]

[1] *Johannes Gutenberg-Universität Mainz, 55099 Mainz, Germany*
[2] *GSI Helmholtzzentrum für Schwerionenforschung GmbH, 64291 Darmstadt, Germany*
[3] *Helmholtz-Institut Mainz, 55099 Mainz, Germany*

**Abstract**

In preparation of gas-phase chemical experiments with moscovium (Mc, element 115), we studied the chemical behavior of the short-lived bismuth radioisotope $^{211}$Bi in helium, argon, and oxygen atmosphere. For that purpose, we performed off-line isothermal gas chromatography experiments at room temperature. Using different carrier gases, the short-lived volatile $^{219}$Rn precursor, provided from an $^{227}$Ac-source, was transported through the Recoil Transfer Chamber (RTC) at the gas-filled separator TASCA and into the mini-**C**ryo-**O**nline **M**ulti detector for **P**hysics **a**nd **C**hemistry of **T**ransactinides (miniCOMPACT) chromatography and detection setup. Internal chromatograms were recorded as a function of various parameters including carrier gas type and flow rate, thus characterizing the novel miniCOMPACT detector array. This aids to optimize the conditions for experiments with superheavy elements. The bismuth progeny of $^{219}$Rn deposited on the SiO$_2$ surface of the miniCOMPACT via diffusion-controlled deposition. Bismuth showed the expected high reactivity towards the SiO$_2$ surface of the miniCOMPACT. Experiments in argon and oxygen atmosphere showed no measurable differences in the deposition distribution of the activity. The intermediate 36-min $^{211}$Pb is a member of the $^{227}$Ac decay chain, feeding the studied bismuth isotope, was taken into account. To extract thermodynamical data from the results, namely the lower limit of the value of the adsorption enthalpy ($-\Delta H_{ads}$) of Bi on SiO$_2$, we performed Monte Carlo simulations, adapted to account for the precursor effect, and compared the experimental results to their output. Simulations were also performed for bismuth's heavier homologue, moscovium, using a theoretically predicted value for $-\Delta H_{ads}$ of this element on SiO$_2$. These suggest moscovium to adsorb in the first part of the miniCOMPACT detection array, in line with recent observations.

## 1. Introduction

The synthesis and characterization of SuperHeavy Elements (SHE) is an interesting interdisciplinary topic of chemistry and physics. [1–3] Through experimental investigations of these artificially produced elements, valuable insights into fundamental questions of chemistry can be gained, also on the validity of trends in chemical behavior in the periodic table of elements towards its heaviest members. Various properties including the electron affinity and the ionization energy exhibit pronounced trends within the groups of the periodic table of elements (PTE). As the atomic number (Z) increases, the influence of relativistic effects on the electronic structure becomes more pronounced, leading to potential deviations in the chemical behavior of SHE compared to that of their lighter homologs. [4,5] Studies of lighter homologs under conditions relevant to studies of SHE are therefore necessary in order to provide a basis against which the behavior of SHEs can be compared.

Theoretical calculations have been employed to investigate the interactions of single SHE atoms with different types of surfaces, enabling predictions for experimental studies and fostering an understanding of their behavior. Results of theoretical calculations of the electronic structure range up to elements with Z=172. [2,6–8] Lately, predictions about the adsorption behaviour of moscovium (Mc, Z=115) and nihonium (Nh, Z=113) on gold and SiO$_2$ surfaces were calculated using fully relativistic periodic density functional theory calculations. [10,11] These calculations suggest both nihonium and moscovium to exhibit higher reactivity towards a SiO$_2$ surface than flerovium [9,12], but lower reactivity than their lighter homologues thallium and bismuth, respectively. The adsorption enthalpy for both nihonium and moscovium is predicted as $-\Delta H_{ads}(\text{Nh}) = -\Delta H_{ads}(\text{Mc}) = 58$ kJ/mol, whereas for thallium and bismuth, it is predicted as $-\Delta H_{ads}(\text{Tl}) = 150.2$ kJ/mol and $-\Delta H_{ads}(\text{Bi}) = 153$ kJ/mol, respectively. [10,13] Experimental chemical data are available for elements or their compounds up to flerovium (Fl, Z=114)

[9,14] and most recently also for moscovium (Mc, Z=115) [15], except for the elements from meitnerium to roentgenium (Z = 109-111).

Single atom gas-phase chromatography is a fast and effective method for investigating the chemistry of SHE and short-lived radioisotopes of lighter homologs. [1,16] Based on the adsorption behavior of single atoms or molecules on different surfaces, conclusions about the volatility and reactivity of the species can be drawn. Liquid-phase chromatography studies have been conducted for elements up to seaborgium (Sg, Z= 106) [17–19], but not yet for heavier elements as only isotopes that have shorter half-lives and lower productions rates are accessible. Adsorption studies by gas-solid chromatography have been carried out for copernicium (Z=112) [20–22] and flerovium. [9,23–25] Initial data for nihonium have been reported in [26], but were not conclusive. [27] Further attempts to study nihonium were unsuccessful. [28,29] Only most recently, nihonium and moscovium were chemically studied, and adsorption enthalpies of $-\Delta H_{ads.}^{SiO_2 \cdot}(Nh) = 58_{-3}^{+8}$ kJ/mol and $-\Delta H_{ads.}^{SiO_2 \cdot}(Mc) = 54_{-5}^{+11}$ kJ/mol were reported, with the data for moscovium being based on four observed atoms only. [15] The main challenge in studying moscovium lies in the short half-life of its most readily available isotope $^{288}$Mc ($t_{1/2}$=195 $_{-1}^{+19}$ ms [30–33]). This requires an even faster chemistry and detection setup compared to previous experiments with longer-lived isotopes of copernicium, nihonium and flerovium. Nihonium and moscovium were shown to be non-volatile and reactive, unlike flerovium and copernicium. Transporting and extracting non-volatile elements like moscovium is particularly challenging, as any contact with a surface before deposition on a detector could result in the deposition and loss of the atom at that position. The demonstration of efficient transport of short-lived bismuth radioisotopes to a fast gas chromatography and detection system is therefore an important step towards optimized moscovium experiments. To provide a basis for the study of the trend from bismuth to moscovium and to validate theoretical predictions, it is crucial to generate benchmark data using bismuth as a lighter homolog of moscovium. Thermochromatography experiments with bismuth on a gold surface found that all species present in this studies were deposited at temperatures above 1100 K, indicating a strong surface interaction. [34] The reported $-\Delta H_{ads}$(Bi) = 269 ± 7 kJ/mol is in good agreement with the theoretical value of $-\Delta H_{ads}$(Bi) = 280 kJ/mol. [10] No volatile species were detected when attempting to produce BiH$_3$ in a cold plasma. [35] Table 1 gives an overview of thermodynamical data from past experiments with a quartz surface and calculations with lead and bismuth on $SiO_2$.

*Table 1: Thermodynamic data for lead and bismuth from several experiments on quartz surfaces, theoretical calculations on $SiO_2$ and empirical relations. $\Delta H_{subl.}$ is the sublimation enthalpy, $-\Delta H_{ads.}^{emp.}$ the adsorption enthalpy determined with an empirical formula, $T_{dep.}$ is the deposition temperature (no adsorption enthalpy reported).*

| Quantity | Pb | Bi | References |
|---|---|---|---|
| $\Delta H_{subl.}$ | 195 kJ/mol | 190 kJ/mol | 36–38 |
| $-\Delta H_{ads.}^{emp.}$ | 161 ± 5 kJ/mol | 157 ± 4 kJ/mol | Calculated after formular in 39 |
| $-\Delta H_{ads.}^{exp.}$ | 142 kJ/mol | 109 kJ/mol | 40 |
| $-\Delta H_{ads.}^{exp.}$ | 207 ± 21 kJ/mol | | 41 |
| $-\Delta H_{ads.}^{exp.}$ | 222 ± 8 kJ/mol | | 42 |
| $-\Delta H_{ads.}^{theo.}$ | 152 kJ/mol | 153 kJ/mol | 10,12 |
| $T_{dep.}$ | 610 ± 40°C | 650 ± 40°C | 43 |

With adsorption enthalphies in the range of 142-222 kJ/mol, both lead and bismuth are expected to be non-volatile at room temperature. Multiple references report similar adsorption behaviour of lead and bismuth, which allows us to formulate an expectation that is later used in the data analysis with Monte Carlo simulations.

## 2. Experimental

In order to demonstrate fast and efficient transport of a non-volatile species, and to provide a basis for comparing the behavior of bismuth and moscovium and to validate theoretical predictions, the interaction strength of bismuth with $SiO_2$ and PTFE (polytetrafluoroethylene) surfaces was investigated in a novel setup. For this series of off-line experiments a $^{227}$Ac ($t_{1/2}$ = 21.77 y, ~50 kBq) emanation source was used. From this source, volatile $^{219}$Rn emanates, cf. Fig. 1. The activity of the source can be described as constant during our studies.

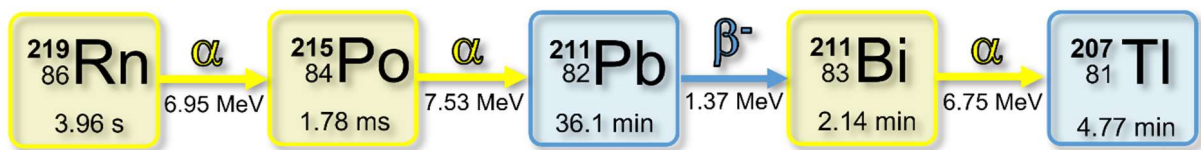

*Figure 1: Decay chain of $^{219}$Rn. The α- and β-decay energies, and half-lives are given for each isotope, yellow: α decay and blue: β⁻ decay.* [44]

Through the carrier gas flow, volatile $^{219}$Rn emanating from the source into the carrier gas is flushed via a 1 m-long PTFE tube (4 mm inner diameter) into the PTFE-coated Recoil Transfer Chamber (RTC, volume $6 \times 4 \times 2$ cm³).[45] The gas loop, as described in [46], provides a steady flow of the carrier gas with user-selected composition through the $^{227}$Ac source, the RTC and the following channel of the miniCOMPACT gas-chromatography and detection setup. Before filling the gas loop, the gases first pass through Hydrosorb® and Oxisorb® cleaning cartridges. The experimental setup is sketched in Fig. 2.

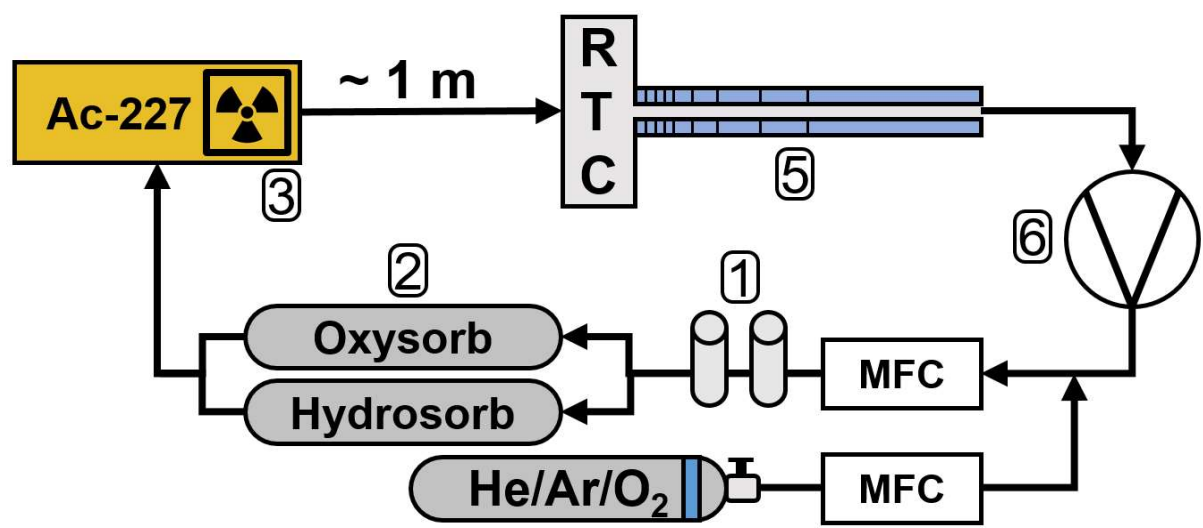

*Figure 2: Schematic of the experimental setup. The gas enters the loop and passes through a mass flow controller (MFC), pressure sensor and dewpoint sensor (1). After passing cleaning cartridges (2), the carrier gas flushes through the $^{227}$Ac emanation source (3) and transports $^{219}$Rn via a 1 m-long PTFE (polytetrafluoroethylene) capillary into the Recoil Transfer Chamber (RTC). Through a $10 \times 0.5$ mm² exit slit, volatile species in the gas phase are extracted from the RTC into the miniCOMPACT (5) which is directly attached to the RTC. The thickness of the RTC backplate at the exit is 0.5 mm. An additional*

*gap of 0.5 mm to the first detector element results in a distance of 1 mm from the gas volume inside the RTC to the first detector, as described in [27]. The carrier gas is circulated with a diaphragm pump (6) within the gas purification system.*

The miniCOMPACT gas-chromatography and detection setup, building up on predecessors described in [27,47], is directly connected to the RTC and comprises two panels, each containing 32 positive-intrinsic-negative (PIN) silicon diodes (thickness: 310 ± 10 μm). Each diode was biased with $U = -20$ V. Each detector panel consists of two consecutive silicon detector arrays, as shown in Figure 3. The first 7 cm-long array contains three groups of different lengths (1, 2 and 4 cm, respectively), with 8 detector elements in each group. The second array is 8 cm long and comprises 8 detector elements with the size of 1×1 cm$^2$ (see Table 2). This special arrangement results in a lower event rate per detector in this part, where all non-volatile species deposited by diffusion-controlled deposition accumulate. In experiments with SHE, the background from unwanted byproducts of the nuclear production reaction can complicate the detection and evaluation of SHE decay chains. The resulting lower rate in the individual detector elements enables statistically significant searches for correlated events separated by a longer correlation time, as the corresponding probability of randomly correlating non-related individual events decreases. The logarithmic detector length distribution does not increase the chromatographic resolution and was thus not providing substantial benefits for the studies reported here.

*Table 2: Dimensions of detector strips.*

| Description of detector strips | Active area including Al contact frame | Inactive gap in between detector elements |
| --- | --- | --- |
| 1 mm | 0.85 × 8.8 mm$^2$ | 0.5 mm |
| 2 mm | 2.1 × 8.7 mm$^2$ | 0.5 mm |
| 5 mm | 4.6 × 8.6 mm$^2$ | 0.5 mm |
| 10 mm | 9.6 × 8.8 mm$^2$ | 0.4 mm |

The diodes are coated with a 30-50 nm layer of silicon dioxide (SiO$_2$), formed through the oxidation of silicon.[48] Two detector panels form a 15 cm-long rectangular channel with a width of 10 mm and a gap between the top and the bottom panel of 0.8 mm, in which the active detector surfaces face each other.[15] The detector signals were processed with the standard data acquisition system of TASCA.[49–51] A photo of the opened miniCOMPACT is shown in Figure 3. The detectors are sensitive to α particles emitted by species inside the channel ($^{219}$Rn, $^{215}$Po, $^{211}$Bi, but not $^{211}$Pb).

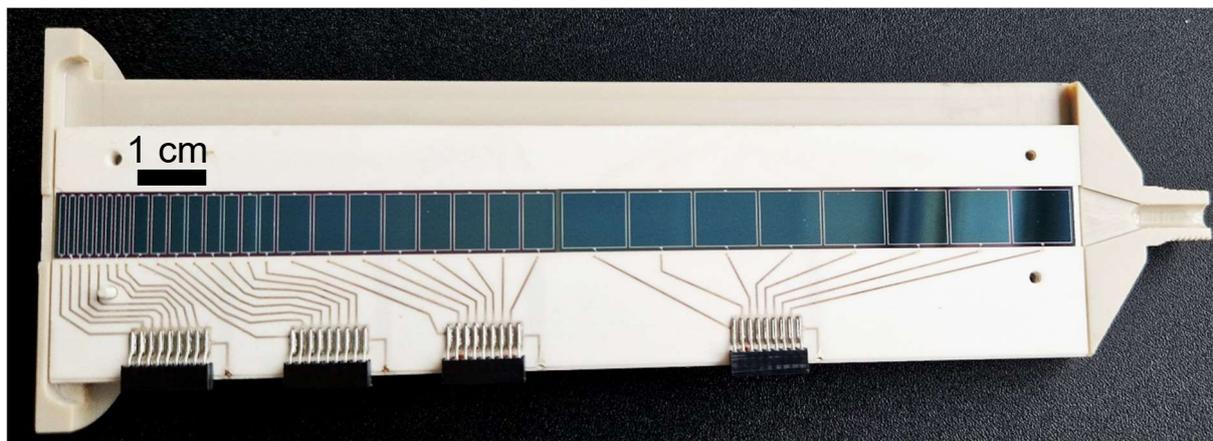

*Figure 3: Picture of one panel of the opened miniCOMPACT detector and gas chromatograph. The left side is directly attached to the RTC.*

The gas loop circulates and cleans the carrier gas in the system. A membrane pump (Pfeiffer Vacuum ACP 40), provides a steady gas flow that is regulated by a mass flow controller (1179 MKS) through the system. Two additional mass flow controllers regulate the inlet of two different pre-dried carrier gases into the loop. The gas composition can be selected using the two mass flow controllers for the inlet of the gases into the system. Helium of 99.9999% purity and argon and oxygen of 99.999% purity were used. The main impurities were nitrogen, oxygen and water. After passing the $^{227}$Ac source, the RTC and the miniCOMPACT, the gas flows through purification cartridges to remove oxygen, water and other impurities down to the levels of a ppm. The gas flow rate and pressure can be monitored and adjusted via software. A detailed description of the gas-loop system can be found in. [52]

The RTC is flushed by the carrier gas, which transports $^{219}$Rn from the source into the RTC and further into the miniCOMPACT detector channel. In a series of experiments, the carrier gas composition (He/Ar/O$_2$) and the gas flow rate (ranging from 1 to 3 L/min) were varied to study how these parameters affect the distribution of $^{211}$Pb and $^{211}$Bi within the column. The flushing and measuring time ranged from a minimum of 61 min to a maximum of 110 min. The carrier gas flow and the data acquisition were started and stopped at the same time. After each experiment, the remaining activity in the detector was left to decay before the next experiment started. This decay without gas flow was then captured in a separate file. The pressure in the RTC was maintained at 1.0 bar, and all experiments were performed at room temperature (20 °C). Chromatographic distributions of $^{219}$Rn and its decay products were measured using α-spectroscopy. Upon reaching the surface, the 36.1-min $^{211}$Pb remains immobilized and will decay via β$^-$-decay, producing the 2.14 min $^{211}$Bi, which will eventually decay via α–particle emission, marking the position in the column. The duration of the time spent in the immobilized state depends on the adsorption enthalpy value of the adsorbed species on the specific substrate material in question.

A fraction of $^{219}$Rn decays as it passes through the RTC. At a gas flow rate of 2 L/min, the volume of the RTC is completely flushed out within 1.44 s (theoretical flush-out of 100% of the volume assuming plug-flow). During this time, about 22% of the radon undergoes two subsequent α-decays. First, $^{219}$Rn decays with a half-life of 3.96 s to $^{215}$Po, which further decays with a half-life of 1.78 ms into $^{211}$Pb (Figure 1). Thus, some $^{211}$Pb is formed inside the RTC and decays to $^{211}$Bi. Due to the long half-live of $^{211}$Pb, the $^{211}$Bi fraction is negligibly small. Among the atoms extracted from the RTC, 78% are extracted as $^{219}$Rn and 22% as $^{211}$Pb, assuming no losses due to adsorption on the walls of the RTC. Due to the extremely short half-life of $^{215}$Po, it is assumed to decay entirely within the RTC directly after the $^{219}$Rn decay and thus $^{215}$Po extraction into the miniCOMPACT system is negligible. In reality though, $^{211}$Pb can encounter the PTFE-coated RTC walls due to diffusion, and can adsorb there as it is non-volatile. This will decrease the yield of $^{211}$Pb compared to $^{219}$Rn, as the noble gas is not retained on the walls. The interaction between $^{219}$Rn and the detector surface inside miniCOMPACT is weak, causing the majority of $^{219}$Rn to exit miniCOMPACT before it decays. Only a small (but quantifiable) fraction undergoes decay-in-flight within miniCOMPACT. The fraction leaving the column was not determined, but estimated using Monte Carlo simulations.

The $^{219}$Rn atoms extracted from the RTC travel through the miniCOMPACT within 27 ms, assuming plug-flow at a gas flow rate of 2 L/min. During this time 0.45% of the atoms decay, according to Monte Carlo simulations assuming $-\Delta H_{ads.}$(Rn) = 20 kJ/mol on SiO$_2$. [53,54] This radon decay-in-flight leads to an evenly distributed activity along the whole chromatography channel, and is thus detected in every detector in the column. At the same time, each decay of $^{219}$Rn inside the RTC yields a 1.78-ms $^{215}$Po daughter atom, which decays almost instantaneously to $^{211}$Pb, which in turn will eventually decay to

$^{211}$Bi. Almost all of the detected $^{211}$Bi is formed after the $^{211}$Pb β$^-$-decay inside the column, at the position that $^{211}$Pb has reached by the time of its decay.

The individual detectors were calibrated using the α-decays of $^{219}$Rn (E$_α$ = 6819 keV) and $^{215}$Po (E$_α$ = 7386 keV).[44] The full width at half maximum (FWHM) of these α lines was 50 keV in pure helium, but varied for different carrier gases. The lines show low-energy tailing, due to the energy losses in the gas, the SiO$_2$ layer and in the dead-layer of the detector.

## 3. Results and discussion

Three measurements in helium, one measurement in argon and one measurement in oxygen as reactive gas have been performed and compared to each other (Table 4). The chromatographic distributions of α activity can vary in steepness, depending on experimental parameters, deposition behavior and nuclear half-life of the species. The spectrum of a representative detector element at the beginning of the column is compared to a spectrum from the end of the column (Figure 4).

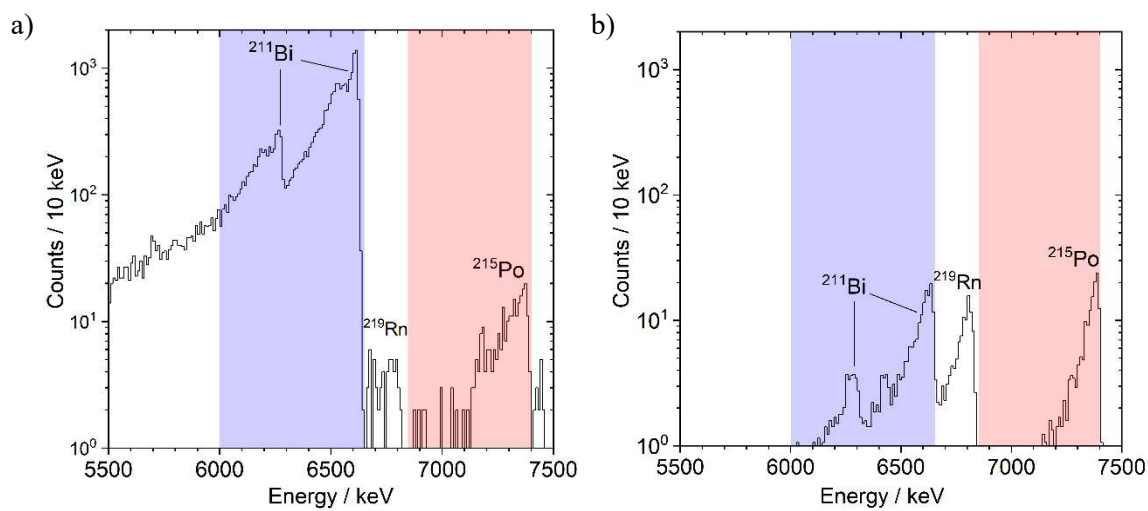

Figure 4: α spectra from detectors 4, 1 mm length (a) and detector 31, 10 mm length (b). The spectra were acquired over 87 min with opened $^{227}$Ac source and helium as carrier gas with 3 L/min gas flow rate at 1 bar pressure. The active area including the contact frame of detector 31 in (b) is 11.3 times larger than that of detector 4 in (a). Thus, the number of counts in (b) was divided by 11.3. The colored areas indicate the areas of integration for the $^{211}$Bi (blue) and the $^{215}$Po (red) peaks.

In the α spectra of each detector element, the $^{219}$Rn peak overlaps with the $^{211}$Bi peak. Therefore, the peak of $^{215}$Po was used to quantify $^{219}$Rn, as the $^{215}$Po decay immediately follows the $^{219}$Rn decay and the $^{215}$Po therefore reproduces the $^{219}$Rn distribute very closely. As can be seen in Figure 4, the $^{215}$Po peaks have the same area at the beginning and at the end of the column, and in fact the $^{215}$Po activity is constant along the whole miniCOMPACT array. This is true for all measurements discussed in this work. The situation is different for $^{211}$Bi. At the beginning of the column, close to the RTC, the intensity of the $^{211}$Bi-peak is much larger than the intensity of the peaks from $^{219}$Rn and $^{215}$Po (Figure 4a). This is an indication that $^{211}$Pb and $^{211}$Bi were extracted from the RTC into the detector channel as non-volatile products of the consecutive α-decays of $^{219}$Rn and $^{215}$Po. As discussed above, the noble gas radon is not retained on the detector surface and decays mostly in-flight. The contribution of the decay in-flight of $^{219}$Rn to the distribution of $^{211}$Bi decay along the column is almost negligible at the beginning of the detector channel, but becomes dominant at the end of the column (Figure 4b), where the peaks originating from $^{219}$Rn, $^{215}$Po and $^{211}$Bi have almost equal intensity. This suggests that a major part of the non-volatile decay products, $^{215}$Po, $^{211}$Pb and $^{211}$Bi, registered at the end of the detector column (Figure 4b), originate from the in-flight decay of $^{219}$Rn rather than from the extraction of those non-volatile elements from the RTC (Figure 5). Carrier gases and volatile elements such as radon interact weakly with the $SiO_2$ detector surface and thus pass the channel almost without retention. On the contrary, reactive species interact with the $SiO_2$ surface more strongly and have significantly longer retention times. Chromatograms were obtained by integrating the α-peaks ($^{215}$Po: 6850-7400 keV; $^{211}$Bi: 6000-6650 keV) in each detector and plotting the integrated number of counts against the column length with a binning of one centimeter. As the errors bars are only of statistical nature, they are relatively small for a high number of counts per centimeter.

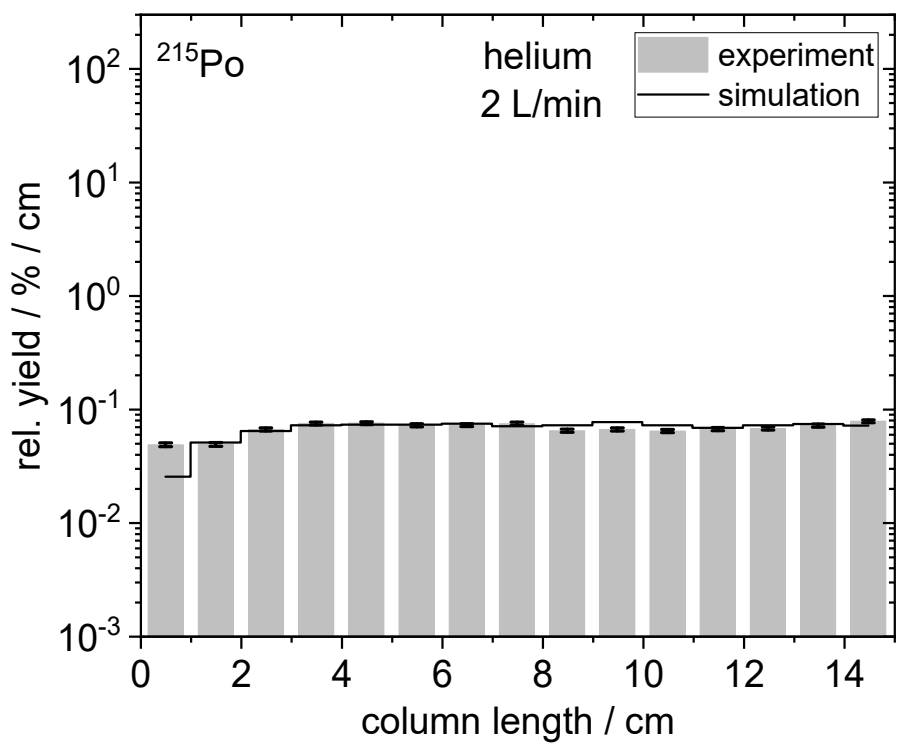

*Figure 5: Distribution pattern of $^{215}$Po (as a direct indication of the $^{219}$Rn decay) on the SiO$_2$-covered miniCOMPACT detector at 1 bar pressure and 2 L/min helium gas flow rate (but representative for all experiments). Grey bars show the experimental relative yield distribution of $^{215}$Po relative to the total bismuth yield with black error bars; the solid line shows the results of a Monte Carlo simulation using $-\Delta H_{ads}(Rn)=20$ kJ/mol and the actual parameters of the experiment.*

The reason for the difference of the Monte Carlo simulation and the experimental data in the first two centimeters lies in the process implemented in the Monte Carlo simulation. The simulated atoms start at the beginning of the column, at 0 cm, in the desorbed state. Then, the mean jump length is calculated and a random jump length is executed, before the first adsorption occurs. Some atoms initially jump further than 1 or 2 cm and have no chance to adsorb and decay in the first two centimeters. This is different in live atoms, which will already have spent a fraction of the corresponding jump before entering the detector array and will on average have a remaining jump length that is shorter than the mean jump length. To disentangle the contribution of $^{211}$Bi entering the miniCOMPACT from the RTC vs. $^{211}$Bi being produced inside the miniCOMPACT, the fraction of $^{211}$Bi activity originating from $^{219}$Rn decay-in-flight in the miniCOMPACT was calculated based on the average $^{215}$Po activity per centimeter (~0.07% of total activity, cf. Figure 5), and this was subtracted from the total measured activity of $^{211}$Bi.

The small contribution of $^{211}$Bi produced from $^{219}$Rn decaying-in-flight within the miniCOMPACT array is insignificant in the first centimeters, but becomes relevant in the last three centimeters, where it dominates the $^{211}$Bi signal and causes a deviation from the exponentially decreasing trend established by the data of the first part of the array.

To determine the extraction efficiency of lead and bismuth from the RTC into the miniCOMPACT, the number of decaying radon atoms within the RTC was calculated at a constant flow rate (assuming plug flow). By comparing Monte Carlo simulations with the experimental in-flight decay of $^{219}$Rn, the total number of radon atoms that passed through the miniCOMPACT during the experiment was estimated. Based on this number, the number of radon decays during the time of the experiment (cf. Table 3) inside the RTC ($N_{Decay,RTC}$(Rn)) was calculated, considering the average time it takes for $^{219}$Rn to pass through the RTC and miniCOMPACT. The extraction efficiency can then be estimated by calculating the ratio of the total number of $^{219}$Rn atoms that decayed in the RTC and the number of $^{211}$Bi decays detected in the miniCOMPACT. The efficiency was determined as 32 ± 6% for the extraction of non-volatile elements from the RTC at the He flow rate of 1 L/min. The extracted atoms were either extracted into the miniCOMPACT without any wall contact or the interaction strength of lead and bismuth on PTFE is insufficient to retain the atoms until they decay inside the RTC. Detailed results for the extraction efficiency for all gases and gas flow rates are shown in Table 3.

Table 3: Calculation of the efficiency of extraction of non-volatile (Pb, Bi) elements from the RTC into the miniCOMPACT in different gases and for different gas-flows rates. $N_{sim.}$(Rn) is the simulated number of radon atoms passing through the RTC and the miniCOMPACT, $N_{Decay,RTC}$(Rn) is the number of atoms that decay during their residence time inside the RTC and $N_{counts}$ is the measured counts of $^{211}$Bi.

| Gas flow rate | He, 1 L/min | He, 2 L/min | He, 3 L/min | Ar, 1 L/min |
| --- | --- | --- | --- | --- |
| $N_{sim.}$(Rn) / $10^6$ | 1.85 ± 0.09 | 2.50 ± 0.08 | 6.30 ± 0.15 | 1.71 ± 0.08 |
| $N_{Decay,RTC}$(Rn) / $10^6$ | 1.21 ± 0.20 | 0.67 ± 0.10 | 1.10 ± 0.14 | 0.98 ± 0.17 |
| $N_{counts}$(Bi) / $10^5$ | 3.94 ± 0.02 | 2.77 ± 0.02 | 5.15 ± 0.03 | 3.45 ± 0.02 |
| Efficiency | 32 ± 6% | 41 ± 6% | 47 ± 6% | 35 ± 6% |

Due to the homogeneous distribution of radon atoms in the RTC volume, the locations of production of non-volatile daugthers of $^{219}$Rn are also homogeneously distributed inside the RTC. In contrast, in online experiments, fusion-evaporation products thermalize within a fraction of the full RTC volume. Thus, the extraction efficiency is potentially even higher. Note, that a plug-flow model was assumed for the calculation of residence time of $^{219}$Rn in the RTC, which might not accurately reflect the real flow conditions in the RTC.

The extracted $^{211}$Pb is detected through the decay of its daughter $^{211}$Bi inside the miniCOMPACT. Figure 6 shows the distributions of $^{211}$Bi, extracted as $^{211}$Pb, on $SiO_2$ in different carrier gases.

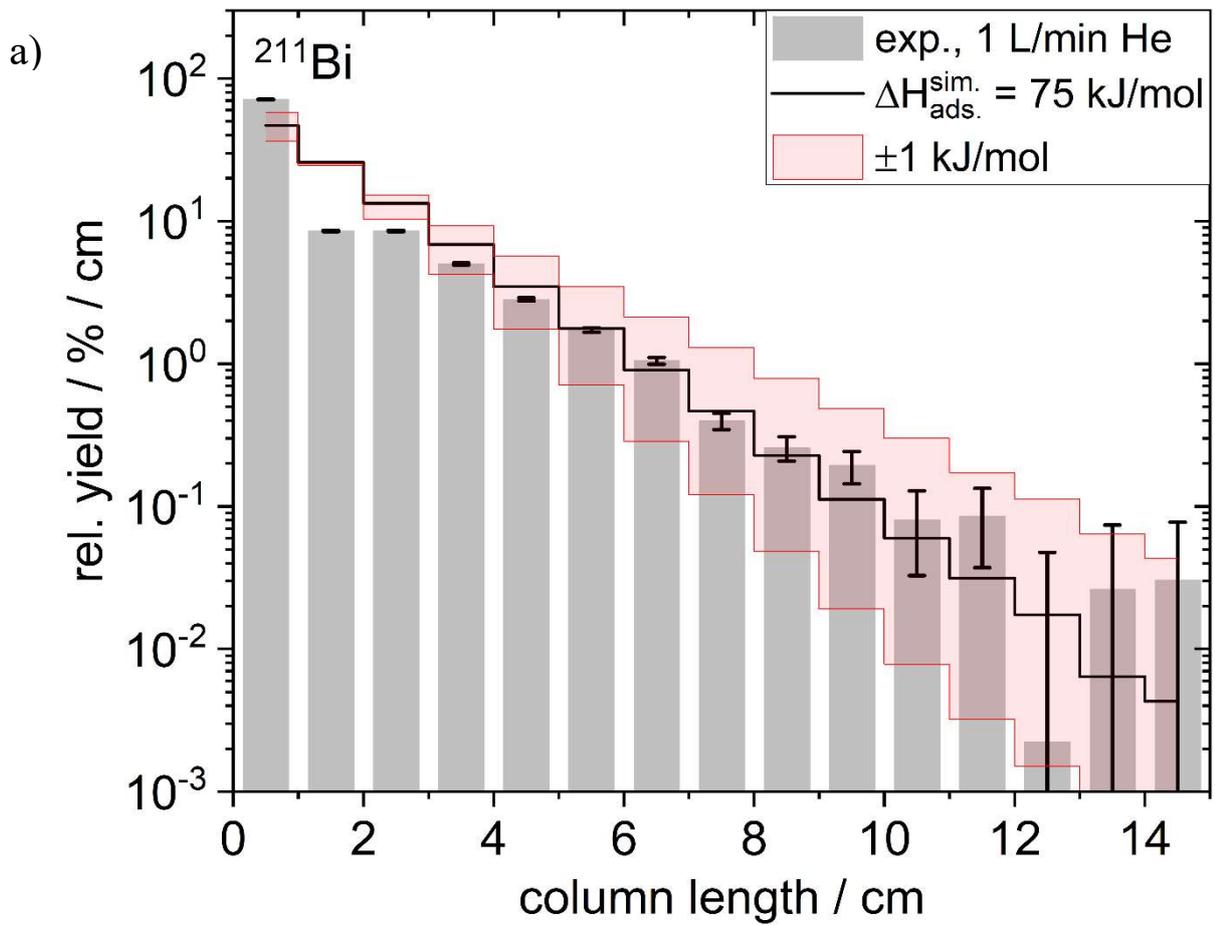

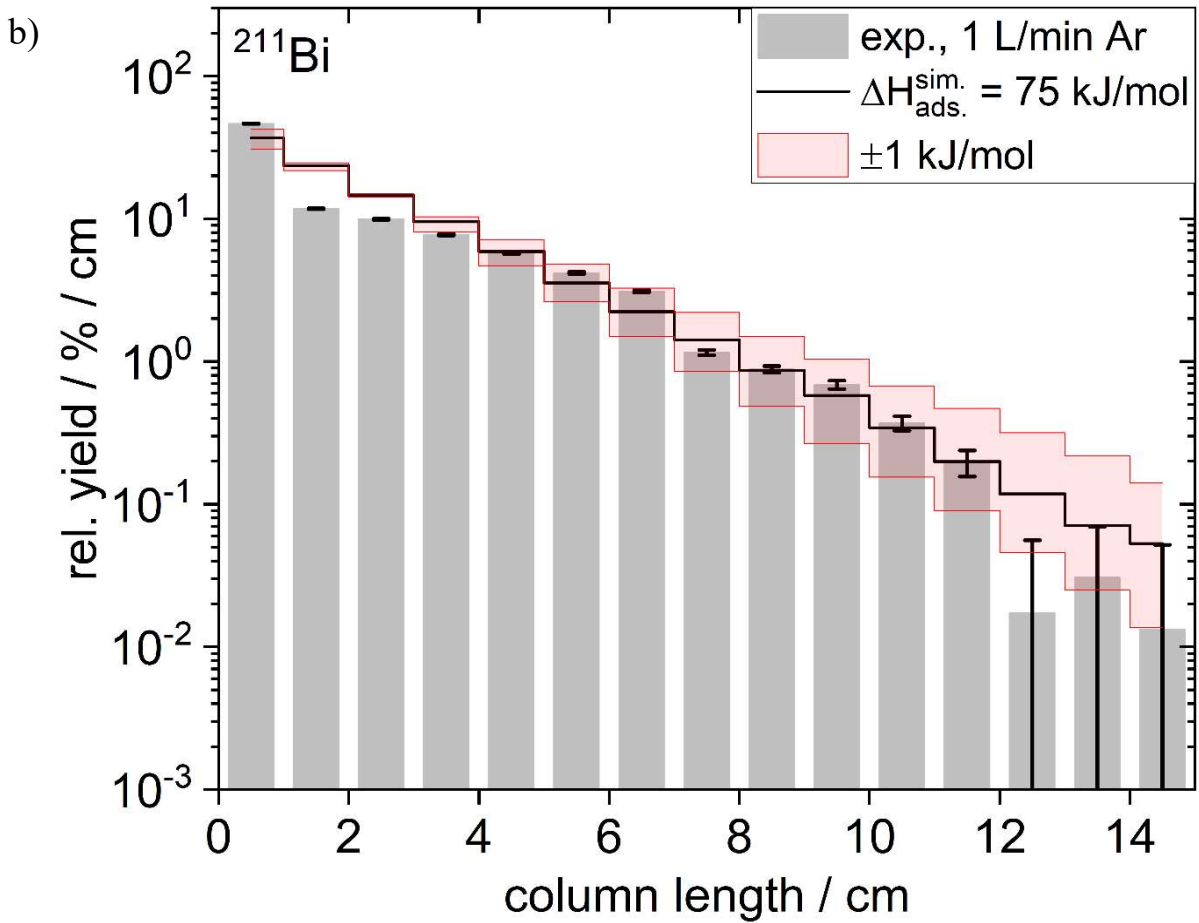

c)

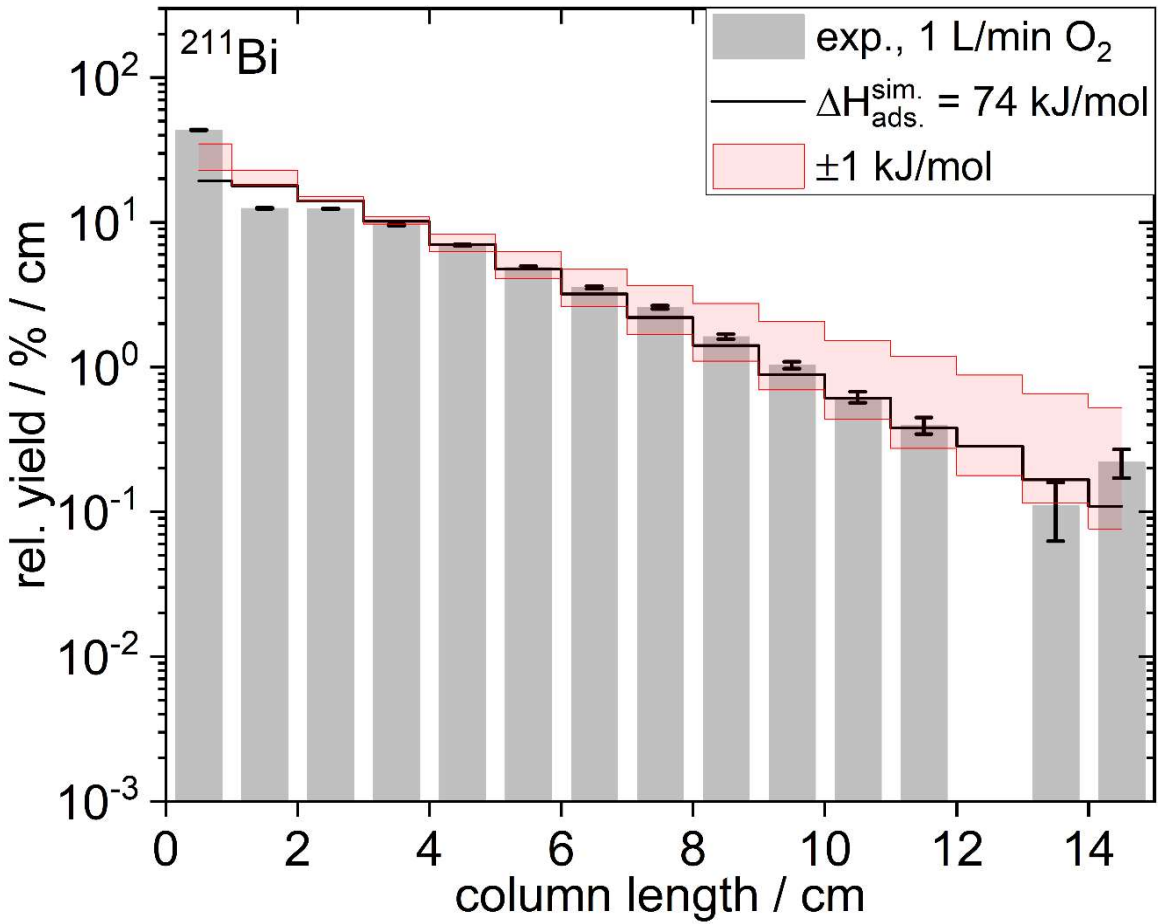

Figure 6: Distribution patterns of $^{211}$Bi, extracted as $^{211}$Pb, on the SiO$_2$-covered miniCOMPACT detector array after subtraction of the $^{211}$Bi activity originating from decay-in-flight inside the channel. Different carrier gases at 1 bar pressure and 1 L/min carrier gas flow rate were used as indicated. Grey bars with black error bars: experimental distributions; black line: results of Monte Carlo simulations (accounting for the precursor effect) with gas parameters identical to the experiments and best agreement to the experimental data measured in detectors 3-15. The red band represents the ± 1 kJ/mol error.

Most (>99%) of the extracted bismuth activity is deposited within the first 8 cm in helium carrier gas, within the first 12 cm in argon, and within the first 11 cm in oxygen with less than 1% being deposited in the remaining length of the column. Besides the carrier gas type, also the gas flow rate influences the distribution. Figure 7 shows the distribution of $^{211}$Bi along the column for different helium flow rates at 1 bar pressure.

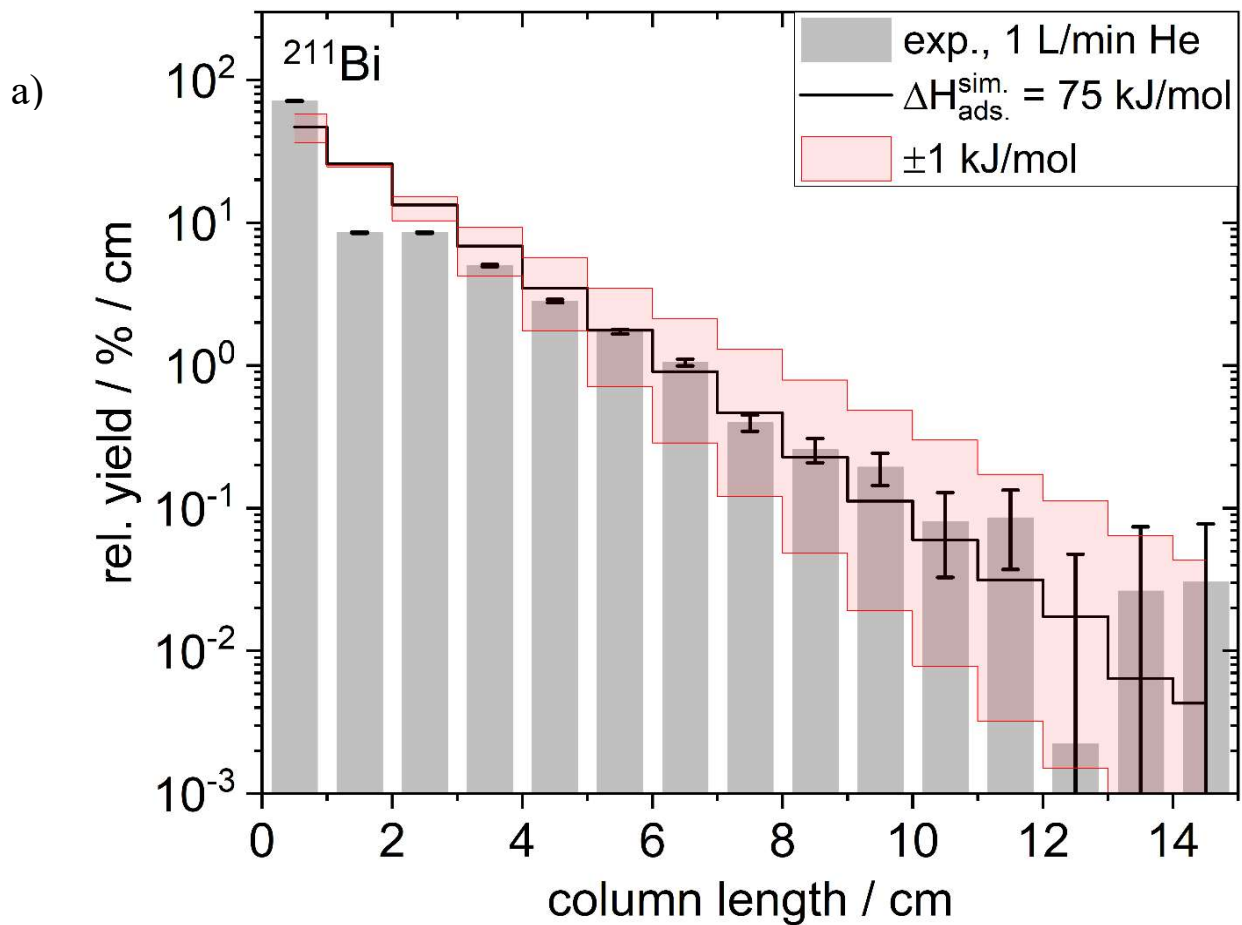
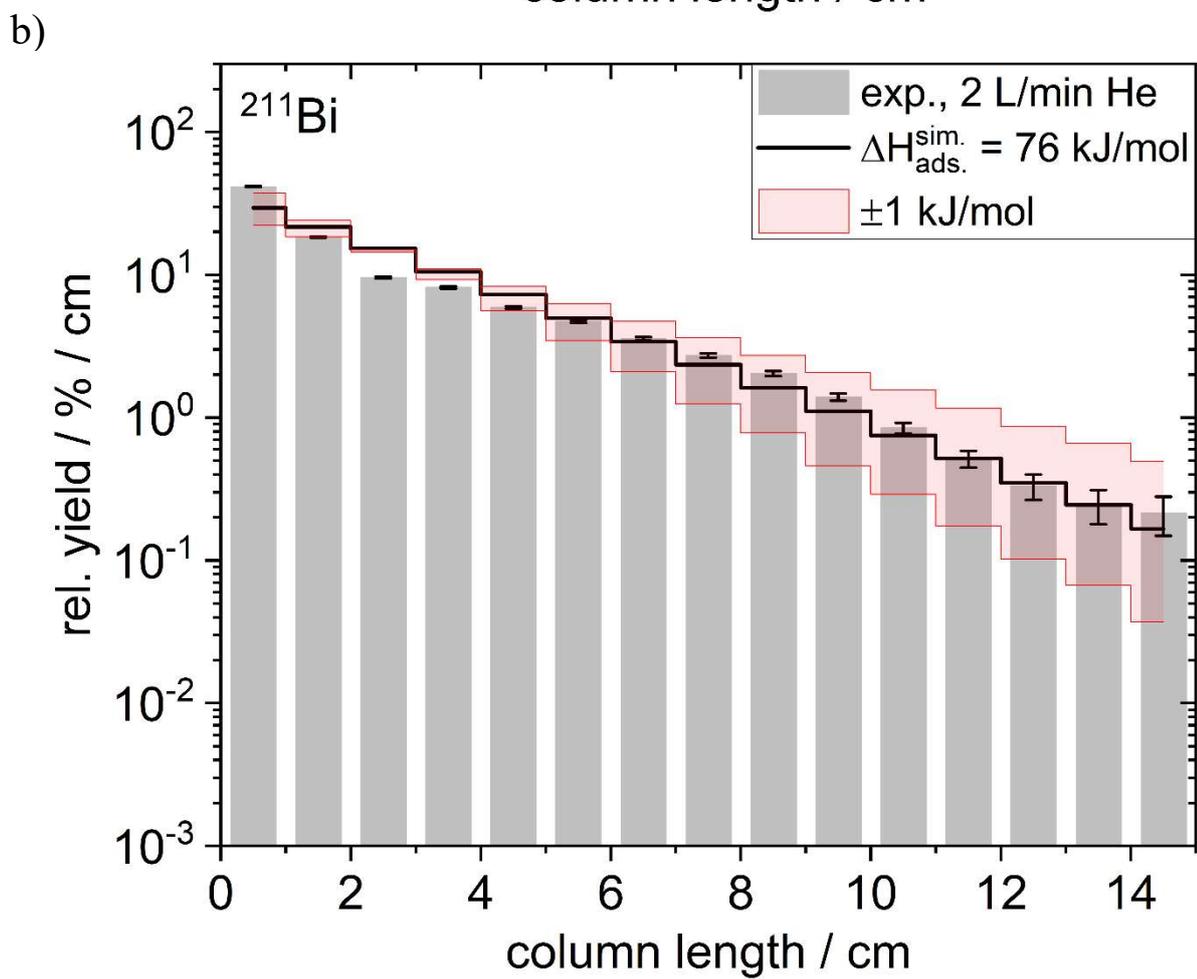

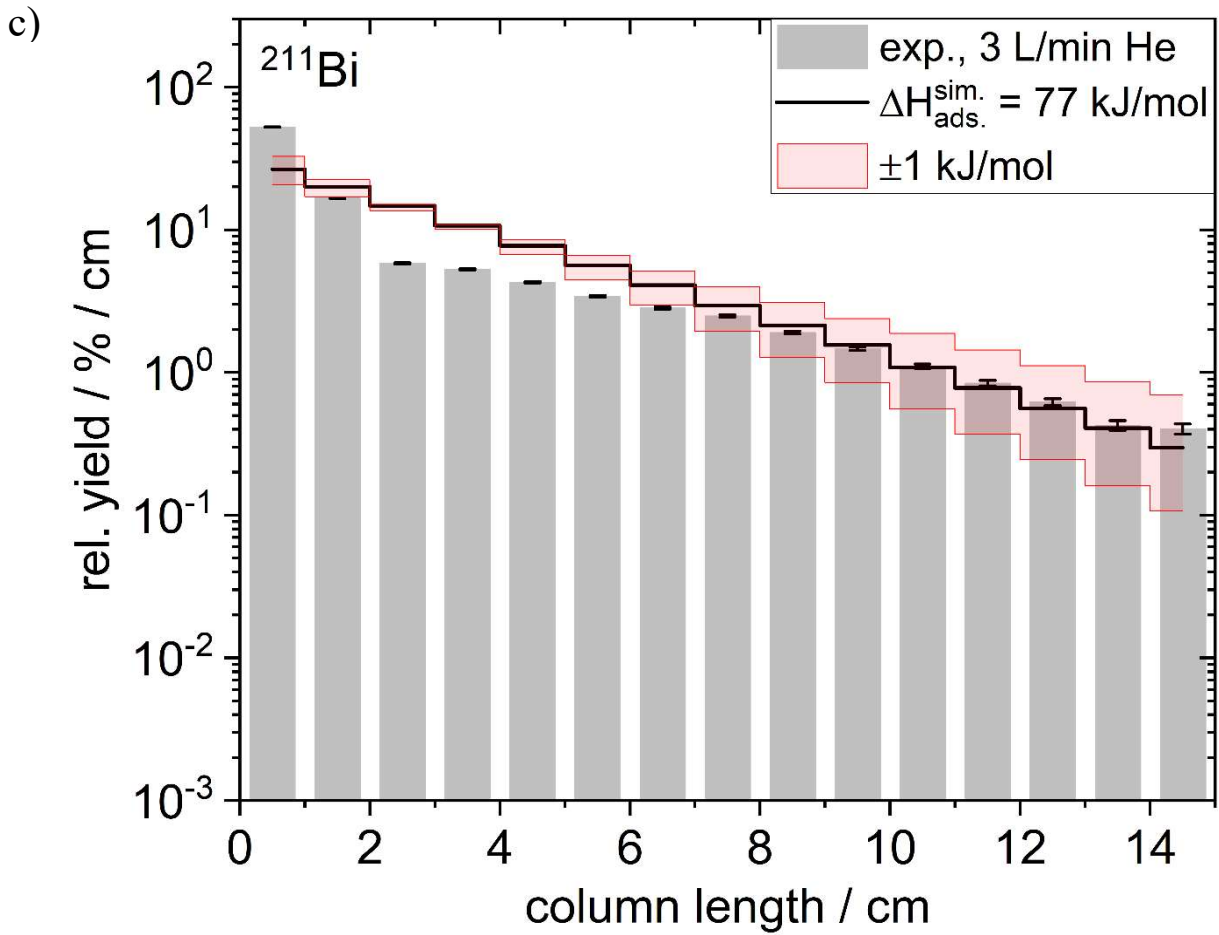

Figure 7: Distribution patterns of $^{211}$Bi on the SiO$_2$ covered miniCOMPACT detector after subtraction of the $^{211}$Bi activity originating from decay-in-flight inside the channel at different helium flow rates (1-3 L/min) at 1 bar pressure. Grey bars with black error bars: experimental distributions; black line: results of Monte Carlo simulations (accounting for the precursor effect) with gas parameters identical to the experiments and best agreement to the experimental data measured in detectors 3-15. The red band represents the ± 1 kJ/mol error.

All distributions show an exponentially decreasing activity along the column. The slope of the distributions decreases with increasing gas flow rate. Most (>99%) of the extracted bismuth is deposited on the first 8 cm at a helium gas flow rate of 1 L/min, with less than 1% being deposited in the remaining length of the column between 8 and 15 cm. For higher gas flow rates, >99% of $^{211}$Bi was deposited on the first 12 cm at 2 L/min and on the first 13 cm at 3 L/min, respectively.

The variations in distributions across various gases and gas flow rates suggest deposition controlled by lateral diffusion, varying as a result of distinct diffusion coefficients of bismuth in the different gases.

Experiments with pure oxygen as the carrier gas aimed to study potential oxide formation of lead/bismuth atoms recoiling after α/β⁻-decay. However, the distribution in pure oxygen (Fig. 6c) showed no significant difference from that in pure argon (Fig. 6b). Likely, also the oxide would deposit due to diffusion-controlled processes. Hence our result cannot provide much information concerning the potential formation of an oxide.

Distributions of $^{215}$Po (Figure 5, decay-in-flight) and $^{211}$Pb/$^{211}$Bi (Figures 6-7, $^{211}$Pb extracted from RTC) were simulated with the Monte Carlo method, as first described in [55]. In this approach, the adsorption enthalpy, $-\Delta H_{ads}$, is the sole input parameter. Monte Carlo simulations used in previous studies simulated only one atomic species. Each atom would start its individual journey through the column at the same position, namely, the beginning of the chromatography channel. The adsorption enthalpy $-\Delta H_{ads}$ was determined by running several simulations with varying $-\Delta H_{ads}$ values and identifying the value where the output best matched the experimental data. A different situation arises when one species is constantly produced from a radioactive precursor that itself travels through the chromatography column. As the mother nucleus (the precursor) decays at a certain position in the column, the daughter nucleus (the progeny) can potentially move further along the detector array until its decay. The likelihood depends on the recoil range of the daughter in the gas atmosphere inside the channel when the mother's decay imparts a recoil away from the detector surface onto the daughter nucleus. The ranges of the recoiling atoms after the $^{211}$Pb β⁻ decay for typical energies of <0.01 keV were calculated using SRIM. [56] The simulations predict a maximum recoil range of 6.1 μm in helium. Due to the low gas velocity in the vicinity of the surface, the probability that a non-volatile atom will be adsorbed at the surface again at a position close to that of its creation is high. Especially recoil emission at shallow angles therefore has a high probability to lead to immediate re-adsorption. Lead as the precursor itself also adsorbs in a characteristic distribution in the column, given the strong interaction with the SiO$_2$ surface [46], in agreement with the theorical value of $-\Delta H_{ads}$ = 152 kJ/mol. [12] Compared to distributions recorded for directly detectable nuclides, this precursor effect can broaden the expected exponential distribution of the activity of a non-volatile daughter and thus, should be considered in the evaluation of the experimental data using Monte Carlo simulations. In the present system, $^{211}$Bi starts adsorbed on the surface and effectively extends the travelling time in the column by the lifetime of the individual bismuth atoms. The mechanism, its impact and a discussion for the case of α-decay is described in detail in [57].

The probabililty for detachment following the adsorption on the surface is very low for reactive elements, because the bond to the surface is strong, leading to long retention times. Thus, the atoms spend most of their lifetime deposited on the surface. In addition, the time for a "long jump" between two cycles of multiple summarized adsorption/desorption steps without significant displacement is relatively short in the rectangular channel of the miniCOMPACT (1.62 ms on average, simulated for $^{211}$Bi at a He flow rate of 2 L/min). If diffusion-controlled deposition occurs, the resulting distribution is determined by the gas composition via the diffusion coefficient of the species in the gas, the pressure, gas flow rate and geometry of the channel. Above a certain value of $-\Delta H_{ads}$, the diffusion-controlled distribution does not become steeper anymore because every (first) surface contact will lead to immobilization for a time longer than the residual nuclear lifetime. If different species have adsorption enthalpy values above this limit, they exhibit identical deposition behavior in our setup. Only a lower limit of the adsorption enthalpy can then be determined. In the simulations, the same value for the adsorption enthalpy was chosen for lead and bismuth, cf. Table 1. If lead were much more reactive than bismuth, it would not influence the distribution, because diffusion controlled deposition occurs already. On the opposite, a volatile precursor would lead to a significantly flatter exponential distribution. [57] If such a volatile precursor decayed randomly in flight, the resulting distribution of $^{211}$Bi would be flat (cf. Figure 5).

Figure 7 compares experimental results of measurements in helium with gas flow rates of 1, 2 and 3 L/min with the Monte Carlo simulations which gave the best agreement for centimeters 3-15 (where we assume a laminar flow regime to exist). The steep, exponentially decreasing distributions confirm the expected strong interaction between lead and bismuth and the $SiO_2$ surface. The diffusion time to the detector surface remains the same for different gas flow rates, but at a higher gas velocity, the same diffusion time to the surface corresponds to a longer travelling inside the channel before a next surface contact occurs. A higher carrier gas flow rate therefore results in a distribution that extends to longer traveled distances along the column. Among measurements at the same flow rates, but in different gases

(Fig. 6) the varying steepness of the distributions arises from a difference in the diffusion coefficient, caused by differences in molar masses (standard values: $M$(He) = 4 g/mol, $M$(Ar) = 40 g/mol, M($O_2$) = 32 g/mol) and densities at the boiling point ($\rho$ (He) = 0.125 g/cm³, $\rho$ (Ar) = 1.3954 g/cm³, $\rho$ ($O_2$) = 1.141 g/cm³) of the carrier gases. [58] The experimental data from 3 - 15 cm are well reproduced by Monte Carlo simulations with an average lower limit based on four measurements (He: 1, 2, 3 L/min, Ar: 1 L/min) of $-\Delta H_{ads}$ = 76 ± 1 kJ/mol for both lead and bismuth as well. Both species of lead and bismuth, i.e., the elemental state as well as oxides, are expected to be non-volatile and to strongly interact with the $SiO_2$ surface. [10,12,46] Thus, the distributions of elemental lead and bismuth and their oxides, if formed, are expected to be similar in the miniCOMPACT. The distribution in Figure 6c points to diffusion-controlled deposition. Therefore, based on the measured distributions, our setup is not sensitive enough to distinguish between the elemental forms and oxides of both lead and bismuth. Table 4 lists the adsorption enthalpy obtained from the Monte Carlo simulation that best fits the experimental data, which is known to represent a limit for diffusion-controlled deposition.

Table 4: Lower limits of the adsorption enthalpies for the ²¹¹Pb/²¹¹Bi system in different gases and at various gas flow rates (rounded to whole numbers). These values are estimated using Monte Carlo simulations that consider the precursor effect. The adsorption enthalpies are obtained from the simulation that best fits the experimental data, representing a lower limit for diffusion-controlled deposition.

| Gas, gas flow rate | collection time | $-\Delta H_{ads}$ (limit) |
|---|---|---|
| He, 1 L/min | 110 min | 75 kJ/mol |
| He, 2 L/min | 61 min | 76 kJ/mol |
| He, 3 L/min | 87 min | 77 kJ/mol |
| Ar, 1 L/min | 93 min | 76 kJ/mol |
| $O_2$, 1 L/min | 82 min | 74 kJ/mol |

The lower limit of the adsorption enthalpy obtained from Monte Carlo simulations that can be discriminated using the miniCOMPACT detector depends on the half-life of the species being studied.

Shorter half-lives lead to correspondingly lower values for this limit, because there is less time available for atoms to make multiple adsorption-desorption cycles before decaying. Therefore, with shorter half-lives, atoms have a reduced probability of traveling further or making additional jumps in the gas stream before they decay. In contrast, studies with isotopes with longer half-lives lead to a higher value for this limit, as atoms have more time to potentially desorb and re-adsorb multiple times, increasing the likelihood of further transport before decay also at higher values of $-\Delta H_{ads}$. It is important to note that the adsorption enthalpy itself is not directly dependent on the half-life of the radionuclide; rather, it is the ability to discriminate between different enthalpy limits that is affected by the half-life. Table 5 provides an overview of the simulated enthalpy limits that can be determined for this setup based on Monte Carlo simulations.

*Table 5: Lower limits of the adsorption enthalpy for a maximum value of $-\Delta H_{ads}$, for which a dependence of the deposition pattern on the $\Delta H_{ads}$ value is observable in our setup, given for half-lives in the range of radioisotopes with 10 ms – 1000 s. Higher values lead to identical deposition patters like that obtained for the limit value. Simulations were performed for a gas flow rate of 2 L/min and using an atomic mass of 288, equal to $^{288}$Mc.*

| $t_{1/2}$ | $-\Delta H_{ads}$ (limit) |
|---|---|
| 1000 s | 84 kJ/mol |
| 10 s | 76 kJ/mol |
| 1 s | 70 kJ/mol |
| 200 ms | 66 kJ/mol |
| 100 ms | 65 kJ/mol |
| 10 ms | 60 kJ/mol |

We note that these simulations lead to steeper predicted deposition patterns compared to the experimental "limit" distribution. In the present case, an element with a known adsorption enthalpy of 150-200 kJ/mol is found as >76 kJ/mol from a simulation that best fits the data, whereas a pure simulation approach would fix the limit value of our setup as >85 kJ/mol. This deviation between the experimental (Table 4) and the simulated limit (Table 5) could be caused by an overestimation of the diffusion of the atoms towards the surface, leading to steeper simulated deposition patterns compared to the experimental deposition. Thus, our determined limit of $-\Delta H_{ads} = 76 \pm 1$ kJ/mol represents a conservative limit, since we cannot account for species with 76 kJ/mol $< -\Delta H_{ads} <$ 85 kJ/mol.

Moscovium and its daughter element nihonium have been predicted to be reactive towards a $SiO_2$ surface with predicted adsorption enthalpies of $-\Delta H_{ads} = 58$ kJ/mol for both elements [10], and recently, first experimental results were published [15] which were in agreement with these predictions. The first members of the decay chain are shown in Figure 8.

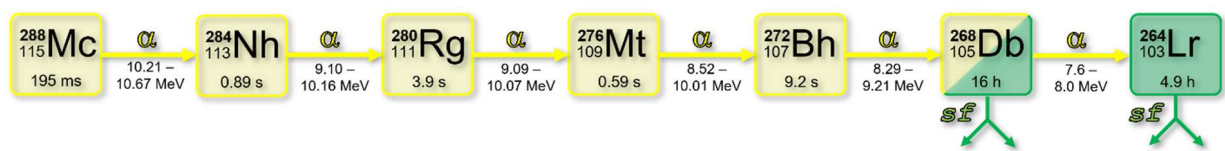

*Figure 8: Decay chain of the short-lived $^{288}$Mc, the heavier homolog of bismuth, yellow: α decay and green: spontaneous fission (sf). [32]*

The described experimental setup has proven suitable for studies of the adsorption behavior of moscovium and nihonium. However, especially in the case of moscovium, the lifetime of the moscovium isotope should be at least comparable to the time needed to flush the RTC volume. Using the setup discussed in this work in on-line experiments with $^{182,183}$Hg, the true extraction time for > 50% of the thermalized atoms is around 500 ms at a gas flow rate of 2 L/min. [27] Thus, for experiments with the short-lived $^{288}$Mc, a gas flow rate of 2 L/min or higher would be needed, since higher gas flow rates correlate with shorter flush-out times. As discussed in [10] the diffusion-controlled deposition of moscovium on the $SiO_2$ surface is expected, leading to the almost complete deposition of $^{288}$Mc in the miniCOMPACT detector channel, with less than 1% of $^{288}$Mc passing through. If moscovium had an interaction strength with $SiO_2$ similar to that of the volatile metal mercury, ($-\Delta H_{ads}$(Hg) = 43 kJ/mol [46]), the retention time in the column at a flow rate of 3 L/min would be only 30 ms. Thus, 87% of $^{288}$Mc ($t_{1/2}=195^{+1}_{-1}$ ms [30–33]) would not decay inside the miniCOMPACT, resulting in a flat distribution from random decay-in-flight of the few retained atoms. In this case, a large fraction of $^{288}$Mc would leave the miniCOMPACT before decay, and thus a second detector with an expectedly more reactive surface like Au, would have to be added. The lower sensitivity limit for $-\Delta H_{ads}$ for diffusion-controlled deposition of $^{288}$Mc is 66 kJ/mol. If $-\Delta H_{ads} >$ 66 kJ/mol, the distribution would not change. The setup is thus sensitive to the determination of a central value of $-\Delta H_{ads}$ for $^{288}$Mc species with $-\Delta H_{ads}$ between ~38 kJ/mol (almost flat distribution) and 66 kJ/mol (diffusion-controlled deposition). Figure 9 shows the predicted fraction of $^{288}$Mc that would deposit inside the miniCOMPACT. Note that the distribution inside the miniCOMPACT depends on $-\Delta H_{ads.}$, allowing for distinction of different interaction strenghts even in a regime where almost 100% are deposited inside, rendering the measured fraction alone insufficient to obtain this information. However, the Monte Carlo simulation uses a high number of atoms, which are not achieveable in currrent experiments. Based on a handful of atoms, the sensitive range must be adjusted to $-\Delta H_{ads}$ between ~46 kJ/mol (~30% deposition inside miniCOMPACT) and 56 kJ/mol (noticeable differences in the distribution inside miniCOMPACT). In the experiment with $^{288}$Mc, the miniCOMPACT acts as a pre-column for non-volatile elements, followed by a gold-coated thermochromatography column like COMPACT to distinguish between volatile species. [15]

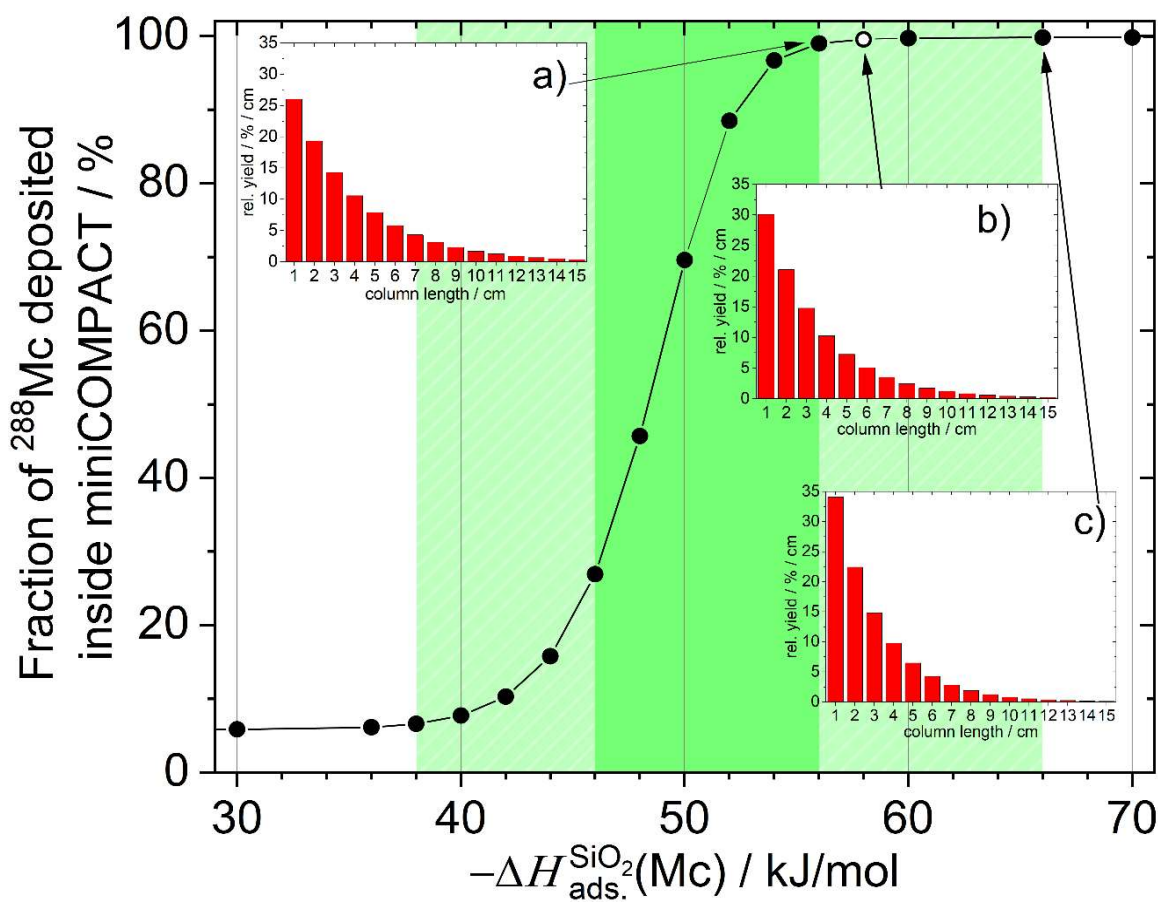

*Figure 9: Monte Carlo simulations yielding the fraction of 195-ms $^{288}$Mc deposited inside miniCOMPACT as a function of adsorption enthalpy. The carrier gas is 3 L/min He:Ar 1:1 at 1 bar.*

*The light green area shows the simulated sensitive range of the miniCOMPACT detector with high statistics. The green area shows the realistic sensitive range in a low statistics experiment. The insets show the deposition pattern for the adsorption enthalpy of 56 kJ/mol (a), where still ~100% is deposited inside the detector. The highlighted data point (open circle) marks the predicted adsorption enthalpy $-\Delta H_{ads}$ = 58 kJ/mol (b). The distribution at the lower limit for diffusion controlled deposition with $-\Delta H_{ads}$ = 66 kJ/mol is shown in (c).*

## 4. Conclusion and outlook

Off-line adsorption studies with the system $^{211}$Pb and $^{211}$Bi were performed. Emanating from a $^{227}$Ac emanation source, $^{219}$Rn and its progenies were flushed through the RTC and the miniCOMPACT chromatography and detection system. The non-volatile $^{211}$Pb formed due to the decay predominantly inside the RTC, while its daughter $^{211}$Bi emerged in the miniCOMPACT. The direct connection of the miniCOMPACT to the RTC allowed the extraction of lead with an estimated efficiency of 32 ± 6% at a gas-flow rate of 1 L/min (cf. Table 3). The evaluated extraction efficiency values for $^{211}$Pb/$^{211}$Bi from the RTC suggest that lead and bismuth can be immobilized on the PTFE surface at room temperature. All species were deposited on the SiO$_2$ surface of the miniCOMPACT in diffusion controlled manner. Monte Carlo simulations, accounting for the precursor effect, were performed and compared with experimental data. In this setup, a lower limit for diffusion controlled adsorption on SiO$_2$ of $-\Delta H_{ads}$ = 76 ± 1 kJ/mol for $^{211}$Pb and $^{211}$Bi was determined. The addition of oxygen did not significantly alter the distributions; differentiating between possible different chemical forms of lead and bismuth was thus not possible. Even with a high gas flow rate of 3 L/min of He, almost 100% of the $^{211}$Pb and $^{211}$Bi activity was deposited inside the 15-cm long detector array. Based on the present results, online experiments with the heavier homolog of bismuth, moscovium ($^{288}$Mc) appeared promising at gas flow rates of > 2 L/min, which lead to fast flush-out and high extraction efficiency. Monte Carlo simulations with this isotope of moscovium suggest that it also should primarily adsorb in the first half of the column due to its reactivity to SiO$_2$ and its relatively short half-life, as it was recently confirmed experimentally. [15] The realistic sensitive range of the miniCOMPACT for the adsorption enthalpy of moscovium based on $^{288}$Mc is ~46-56 kJ/mol.

**Funding information**

D. Dietzel acknowledges the Deutschlandstipendium scholarship. We acknowledge funding from the German BMBF (contract Nr. 05P21UMFN2).

**References**

1. Schädel, M., Shaughnessy, D., Eds. *The Chemistry of Superheavy Elements*, 2. ed.; Springer: Berlin, 2014.

2. Schwerdtfeger, P.; Smits, O. R.; Pyykkö, P. The Periodic Table and the Physics that Drives it. *Nat. Rev. Chem.* **2020,** *4* (7), pp. 359–380. DOI: 10.1038/s41570-020-0195-y.

3. Smits, O. R.; Düllmann, Ch. E.; Indelicato, P.; Nazarewicz, W.; Schwerdtfeger, P. The Quest for Superheavy Elements and the Limit of the Periodic Table. *Nat. Rev. Phys.* **2024,** *6*, pp. 86–98. DOI: 10.1038/s42254-023-00668-y.

4. Pyykkö, P. Relativistic Effects in Structural Chemistry. *Chem. Rev.* **1988,** *88* (3), pp. 563–594. DOI: 10.1021/cr00085a006.

5. Pyykkö, P. Relativistic Effects in Chemistry: More Common Than You Thought. *Annu. Rev. Phys. Chem.* **2012,** *63*, pp. 45–64. DOI: 10.1146/annurev-physchem-032511-143755.

6. Fricke, B.; Greiner, W.; Waber, J. T. The Continuation of the Periodic Table up to Z = 172. The Chemistry of Superheavy Elements. *Theoret. Chim. Acta* **1971,** *21* (3), pp. 235–260. DOI: 10.1007/BF01172015.

7. Pyykkö, P. A Suggested Periodic Table up to Z≤172, Based on Dirac-Fock Calculations on Atoms and Ions. *Phys. Chem. Chem. Phys.* **2011,** *13* (1), pp. 161–168. DOI: 10.1039/c0cp01575j.

8. Fricke, B.; McMinn, J. Chemical and physical properties of superheavy elements. *Naturwissenschaften* **1976,** *63* (4), pp. 162–170. DOI: 10.1007/BF00624214.

9. Yakushev, A.; Lens, L.; Düllmann, Ch. E.; Khuyagbaatar, J.; Jäger, E.; Krier, J.; Runke, J.; Albers, H. M.; Asai, M.; Block, M.; Despotopulos, J.; Di Nitto, A.; Eberhardt, K.; Forsberg, U.; Golubev, P.; Götz, M.; Götz, S.; Haba, H.; Harkness-Brennan, L.; Herzberg, R.-D.; Heßberger, F. P.; Hinde, D.; Hübner, A.; Judson, D.; Kindler, B.; Komori, Y.; Konki, J.; Kratz, J. V.; Kurz, N.; Laatiaoui, M.; Lahiri, S.; Lommel, B.; Maiti, M.; Mistry, A. K.; Mokry, C.; Moody, K. J.; Nagame, Y.; Omtvedt, J. P.; Papadakis, P.; Pershina, V.; Rudolph, D.; Samiento, L. G.; Sato, T. K.; Schädel, M.; Scharrer, P.; Schausten, B.; Shaughnessy, D. A.; Steiner, J.; Thörle-Pospiech, P.; Toyoshima, A.; Trautmann, N.; Tsukada, K.; Uusitalo, J.; Voss, K.-O.; Ward, A.; Wegrzecki, M.; Wiehl, N.; Williams, E.; Yakusheva, V. On the Adsorption and Reactivity of Element 114, Flerovium. *Front. Chem.* **2022,** *10*, pp. 976635. DOI: 10.3389/fchem.2022.976635.

10. Pershina, V.; Iliaš, M.; Yakushev, A. Reactivity of the Superheavy Element 115, Mc, and Its Lighter Homologue, Bi, with Respect to Gold and Hydroxylated Quartz Surfaces from Periodic Relativistic DFT Calculations: A Comparison with Element 113, Nh. *Inorg. Chem.* **2021,** *60* (13), pp. 9796–9804. DOI: 10.1021/acs.inorgchem.1c01076.

11. Trombach, L.; Ehlert, S.; Grimme, S.; Schwerdtfeger, P.; Mewes, J.-M. Exploring the Chemical Nature of Super-Heavy Main-Group Elements by Means of Efficient Plane-Wave Density-Functional Theory. *Phys. Chem. Chem. Phys.* **2019,** *21* (33), pp. 18048–18058. DOI: 10.1039/c9cp02455g.

12. Pershina, V. A Relativistic Periodic DFT Study on Interaction of Superheavy Elements 112 (Cn) and 114 (Fl) and Their Homologs Hg and Pb, Respectively, with a Quartz Surface. *Phys. Chem. Chem. Phys.* **2016,** *18* (26), pp. 17750–17756. DOI: 10.1039/C6CP02253G.

13. Pershina, V. A Theoretical Study on the Adsorption Behavior of Element 113 and Its Homologue Tl on a Quartz Surface: Relativistic Periodic DFT Calculations. *J. Phys. Chem. C* **2016,** *120* (36), pp. 20232–20238. DOI: 10.1021/acs.jpcc.6b07834.

14. Türler, A.; Pershina, V. Advances in the Production and Chemistry of the Heaviest Elements. *Chem. Rev.* **2013,** *113* (2), pp. 1237–1312. DOI: 10.1021/cr3002438.

15. Yakushev, A.; Khuyagbaatar, J.; Düllmann, Ch. E.; Block, M.; Cantermir, R. A.; Cox, D. M.; Dietzel, D.; Giacoppo, F.; Hrabar, Y.; Iliaš, M.; Jäger, E.; Krier, J.; Krupp, D.; Kurz, N.; Lens, L.; Löchner, S.; Mokry, C.; Mošat, P.; Pershina, V.; Raeder, S.; Rudolph, D.; Runke, J.; Sarmiento, L. G.; Schausten, B.; Scherer, U.; Thörle-Pospiech, P.; Trautmann, N.; Wegrzecki, M.; Wieczorek, P. Manifestation of relativistic effects in the chemical properties of nihonium and moscovium revealed by gas chromatography studies. *Front. Chem.* **2024** (12), pp. 1474820. DOI: 10.3389/fchem.2024.1474820.

16. Zvára, I. *The Inorganic Radiochemistry of Heavy Elements*: *Methods for studying gaseous compounds;* Springer: New York, NY, 2008.

17. Nagame, Y.; Kratz, J. V.; Schädel, M. Chemical Studies of Elements with Z≥104 in Liquid Phase. *Nucl. Phys. A* **2015,** *944*, pp. 614–639. DOI: 10.1016/j.nuclphysa.2015.07.013.

18. Schädel, M.; Brüchle, W.; Dressler, R.; Eichler, B.; Gäggeler, H. W.; Günther, R.; Gregorich, K. E.; Hoffman, D. C.; Hübener, S.; Jost, D. T.; Kratz, J. V.; Paulus, W.; Schumann, D.; Timokhin, S.; Trautmann, N.; Türler, A.; Wirth, G.; Yakushev, A. Chemical Properties of Element 106 (seaborgium). *Nature* **1997,** *388* (6637), pp. 55–57. DOI: 10.1038/40375.

19. Schädel, M.; Brüchle, W.; Schausten, B.; Schimpf, E.; Jäger, E.; Wirth, G.; Günther, R.; Kratz, J. V.; Paulus, W.; Seibert, A.; Thörle, P.; Trautmann, N.; Zauner, S.; Schumann, D.; Andrassy, M.; Misiak, R.; Gregorich, K. E.; Hoffman, D. C.; Lee, D. M.; Sylwester, E. R.; Nagame, Y.; Oura, Y. First Aqueous Chemistry with Seaborgium (Element 106). *Radiochim. Acta* **1997,** *77* (3), pp. 149–160. DOI: 10.1524/ract.1997.77.3.149.

20. Eichler, R.; Aksenov, N. V.; Belozerov, A. V.; Bozhikov, G. A.; Chepigin, V. I.; Dmitriev, S. N.; Dressler, R.; Gäggeler, H. W.; Gorshkov, V. A.; Haenssler, F.; Itkis, M. G.; Laube, A.; Lebedev, V. Y.; Malyshev, O. N.; Oganessian, Y. T.; Petrushkin, O. V.; Piguet, D.; Rasmussen, P.; Shishkin, S. V.; Shutov, A. V.; Svirikhin, A. I.; Tereshatov, E. E.; Vostokin, G. K.; Wegrzecki, M.; Yeremin, A. V. Chemical Characterization of Element 112. *Nature* **2007,** *447* (7140), pp. 72–75. DOI: 10.1038/nature05761.

21. Eichler, R.; Aksenov, N. V.; Belozerov, A. V.; Bozhikov, G. A.; Chepigin, V. I.; Dmitriev, S. N.; Dressler, R.; Gäggeler, H. W.; Gorshkov, A. V.; Itkis, M. G.; Haenssler, F.; Laube, A.; Lebedev, V. Y.; Malyshev, O. N.; Oganessian, Y. T.; Petrushkin, O. V.; Piguet, D.; Popeko, A. G.; Rasmussen, P.; Shishkin, S. V.; Serov, A. A.; Shutov, A. V.; Svirikhin, A. I.; Tereshatov, E. E.; Vostokin, G. K.; Wegrzecki, M.; Yeremin, A. V. Thermochemical and Physical Properties of Element 112. *Angew. Chem. Int. Ed.* **2008,** *47* (17), pp. 3262–3266. DOI: 10.1002/anie.200705019.

22. S. Soverna. Attempt to Chemically Characterize Element 112: Inauguraldissertation der Philosophisch-naturwissenschaftlichen Fakultät der Universität Bern, Bern, Switzerland, 2004.

23. Eichler, R.; Aksenov, N. V.; Albin, Y. V.; Belozerov, A. V.; Bozhikov, G. A.; Chepigin, V. I.; Dmitriev, S. N.; Dressler, R.; Gäggeler, H. W.; Gorshkov, V. A.; Henderson, G. S.; al, e. Indication for a Volatile Element 114. *Radiochim. Acta* **2010,** *98* (3). DOI: 10.1524/ract.2010.1705.

24. Yakushev, A.; Gates, J. M.; Türler, A.; Schädel, M.; Düllmann, Ch. E.; Ackermann, D.; Andersson, L.-L.; Block, M.; Brüchle, W.; Dvorak, J.; Eberhardt, K.; Essel, H. G.; Even, J.; Forsberg, U.; Gorshkov, A.; Graeger, R.; Gregorich, K. E.; Hartmann, W.; Herzberg, R.-D.;

Hessberger, F. P.; Hild, D.; Hübner, A.; Jäger, E.; Khuyagbaatar, J.; Kindler, B.; Kratz, J. V.; Krier, J.; Kurz, N.; Lommel, B.; Niewisch, L. J.; Nitsche, H.; Omtvedt, J. P.; Parr, E.; Qin, Z.; Rudolph, D.; Runke, J.; Schausten, B.; Schimpf, E.; Semchenkov, A.; Steiner, J.; Thörle-Pospiech, P.; Uusitalo, J.; Wegrzecki, M.; Wiehl, N. Superheavy Element Flerovium (Element 114) Is a Volatile Metal. *Inorg. Chem.* **2014,** *53* (3), pp. 1624–1629. DOI: 10.1021/ic4026766.

25. Yakushev, A.; Eichler, R. Gas-phase Chemistry of Element 114, Flerovium. *EPJ Web Conf.* **2016,** *131*, pp. 7003. DOI: 10.1051/epjconf/201613107003.

26. Dmitriev, S. N.; Aksenov, N. V.; Albin, Y. V.; Bozhikov, G. A.; Chelnokov, M. L.; Chepygin, V. I.; Eichler, R.; Isaev, A. V.; Katrasev, D. E.; Lebedev, V. Y.; Malyshev, O. N.; Petrushkin, O. V.; Porobanuk, L. S.; Ryabinin, M. A.; Sabel'nikov, A. V.; Sokol, E. A.; Svirikhin, A. V.; Starodub, G. Y.; Usoltsev, I.; Vostokin, G. K.; Yeremin, A. V. Pioneering Experiments on the Chemical Properties of Element 113. *Mendeleev Commun.* **2014,** *24* (5), pp. 253–256. DOI: 10.1016/j.mencom.2014.09.001.

27. Yakushev, A.; Lens, L.; Düllmann, Ch. E.; Block, M.; Brand, H.; Calverley, T.; Dasgupta, M.; Di Nitto, A.; Götz, M.; Götz, S.; Haba, H.; Harkness-Brennan, L.; Herzberg, R.-D.; Heßberger, F. P.; Hinde, D.; Hübner, A.; Jäger, E.; Judson, D.; Khuyagbaatar, J.; Kindler, B.; Komori, Y.; Konki, J.; Kratz, J. V.; Krier, J.; Kurz, N.; Laatiaoui, M.; Lommel, B.; Lorenz, C.; Maiti, M.; Mistry, A. K.; Mokry, C.; Nagame, Y.; Papadakis, P.; Såmark-Roth, A.; Rudolph, D.; Runke, J.; Sarmiento, L. G.; Sato, T. K.; Schädel, M.; Scharrer, P.; Schausten, B.; Steiner, J.; Thörle-Pospiech, P.; Toyoshima, A.; Trautmann, N.; Uusitalo, J.; Ward, A.; Wegrzecki, M.; Yakusheva, V. First Study on Nihonium (Nh, Element 113) Chemistry at TASCA. *Front. Chem.* **2021,** *9*, pp. 753738. DOI: 10.3389/fchem.2021.753738.

28. Türler, A.; Eichler, R.; Yakushev, A. Chemical Studies of Elements with Z≥104 in Gas Phase. *Nucl. Phys. A* **2015,** *944*, pp. 640–689. DOI: 10.1016/j.nuclphysa.2015.09.012.

29. Aksenov, N. V.; Steinegger, P.; Abdullin, F. S.; Albin, Y. V.; Bozhikov, G. A.; Chepigin, V. I.; Eichler, R.; Lebedev, V. Y.; Madumarov, A. S.; Malyshev, O. N.; Petrushkin, O. V.; Polyakov, A. N.; Popov, Y. A.; Sabel'nikov, A. V.; Sagaidak, R. N.; Shirokovsky, I. V.; Shumeiko, M. V.; Starodub, G. Y.; Tsyganov, Y. S.; Utyonkov, V. K.; Voinov, A. A.; Vostokin, G. K.; Yeremin, A. V.; Dmitriev, S. N. On the Volatility of Nihonium (Nh, Z = 113). *Eur. Phys. J. A* **2017,** *53*. DOI: 10.1140/epja/i2017-12348-8.

30. Rudolph, D.; Sarmiento, L. G.; Forsberg, U. Nuclear Structure Notes on Element 115 Decay Chains. In *Nuclear Structure Notes on Element 115 Decay Chains*; AIP Publishing LLC, 2015, p 30015. DOI: 10.1063/1.4932259.

31. Oganessian, Y. T.; Utyonkov, V. K. Super-heavy Element Research. *Rep. Prog. Phys. (Reports on progress in physics. Physical Society (Great Britain))* **2015,** *78* (3), pp. 36301. DOI: 10.1088/0034-4885/78/3/036301.

32. Oganessian, Y. T.; Utyonkov, V. K.; Kovrizhnykh, N. D.; Abdullin, F. S.; Dmitriev, S. N.; Ibadullayev, D.; Itkis, M. G.; Kuznetsov, D. A.; Petrushkin, O. V.; Podshibiakin, A. V.; Polyakov, A. N.; Popeko, A. G.; Sagaidak, R. N.; Schlattauer, L.; Shirokovski, I. V.; v. d. Shubin; Shumeiko, M. V.; Solovyev, D. I.; Tsyganov, Y. S.; Voinov, A. A.; Subbotin, V. G.; Bodrov, A. Y.; Sabel'nikov, A. V.; Khalkin, A. V.; Zlokazov, V. B.; Rykaczewski, K. P.; King, T. T.; Roberto, J. B.; Brewer, N. T.; Grzywacz, R. K.; Gan, Z. G.; Zhang, Z. Y.; Huang, M. H.; Yang, H. B. First experiment at the Super Heavy Element Factory: High cross section of $^{288}$Mc in the $^{243}$Am+$^{48}$Ca reaction and identification of the new isotope $^{264}$Lr. *Phys. Rev. C* **2022,** *106* (3), L031301. DOI: 10.1103/PhysRevC.106.L031301.

33. Oganessian, Y. T.; Utyonkov, V. K.; Kovrizhnykh, N. D.; Abdullin, F. S.; Dmitriev, S. N.; Dzhioev, A. A.; Ibadullayev, D.; Itkis, M. G.; Karpov, A. V.; Kuznetsov, D. A.; Petrushkin, O. V.; Podshibiakin, A. V.; Polyakov, A. N.; Popeko, A. G.; Rogov, I. S.; Sagaidak, R. N.;

Schlattauer, L.; v. d. Shubin; Shumeiko, M. V.; Solovyev, D. I.; Tsyganov, Y. S.; Voinov, A. A.; Subbotin, V. G.; Bodrov, A. Y.; Sabel'nikov, A. V.; Khalkin, A. V.; Rykaczewski, K. P.; King, T. T.; Roberto, J. B.; Brewer, N. T.; Grzywacz, R. K.; Gan, Z. G.; Zhang, Z. Y.; Huang, M. H.; Yang, H. B. New isotope $^{286}$Mc produced in the $^{243}$Am+$^{48}$Ca reaction. *Phys. Rev. C* **2022,** *106* (6). DOI: 10.1103/PhysRevC.106.064306.

34. Maugeri, E. A.; Neuhausen, J.; Eichler, R.; Dressler, R.; Rijpstra, K.; Cottenier, S.; Piguet, D.; Vögele, A.; Schumann, D. Adsorption of Volatile Polonium and Bismuth Species on Metals in Various Gas Atmospheres: Part I – Adsorption of Volatile Polonium and Bismuth on Gold. *Radiochim. Acta* **2016,** *104* (11), pp. 757–767. DOI: 10.1515/ract-2016-2573.

35. Tiebel, G.; Eichler, R.; Steinegger, P. *PSI LRC Annual Reports 2019*: *In Situ Production of BiH$_3$ in Cold Plasmas;* Villigen, Switzerland, 2020, pp. 11–12. https://www.psi.ch/en/lrc/annual-reports (accessed 01.10.2024).

36. Oxtoby, D. W.; Gillis, H. P.; Butler, L. J. *Principles of modern chemistry*, 8. ed.; Cengage Learning: Andover, 2016.

37. Archer, D. G. Enthalpy of Fusion of Bismuth: A Certified Reference Material for Differential Scanning Calorimetry. *J. Chem. Eng. Data* **2004,** *49* (5), pp. 1364–1367. DOI: 10.1021/je049913p.

38. Zhang, Y.; Evans, J. R. G.; Yang, S. Corrected Values for Boiling Points and Enthalpies of Vaporization of Elements in Handbooks. *J. Chem. Eng. Data* **2011,** *56* (2), pp. 328–337. DOI: 10.1021/je1011086.

39. Steinegger, P.; Asai, M.; Dressler, R.; Eichler, R.; Kaneya, Y.; Mitsukai, A.; Nagame, Y.; Piguet, D.; Sato, T. K.; Schädel, M.; Takeda, S.; Toyoshima, A.; Tsukada, K.; Türler, A.; Vascon, A. Vacuum Chromatography of Tl on SiO$_2$ at the Single-Atom Level. *J. Phys. Chem. C* **2016,** *120* (13), pp. 7122–7132. DOI: 10.1021/acs.jpcc.5b12033.

40. Gäggeler, H. W.; Eichler, B.; Greulich, N.; Herrmann, G.; Trautmann, N. Vacuum-Thermochromatography of Carrier-free Species. *Radiochim. Acta* **1986,** *40* (3), pp. 137–144. DOI: 10.1524/ract.1986.40.3.137.

41. Haenssler, F.; Eichler, R.; Gäggeler, H. W.; Dressler, R.; Piguet, D.; Schnippering, M. *PSI LRC Annual Report 2004*: *Thermochromatographic investiagation of $^{212}$Pb on quartz;* Villigen, Switzerland, 2005.

42. Fan, W.; Gäggeler, H. Thermochromatography of Carrier-free Lead in Quartz Columns with Hydrogen and Argon as Carrier Gases. *Radiochim. Acta* **1982,** *31* (1-2), pp. 95–98. DOI: 10.1524/ract.1982.31.12.95.

43. B. Eichler. Исследование распределения некоторых продуктов ядерных реакций без носителей методом термохроматографии в потоке водорода [Investigation of the Distribution of Certain Products of Nuclear Reactions without Carriers Using the Method of Thermochromatography in a Hydrogen Stream]. *Preprint JINR-P12-6662, Joint Institute for Nuclear Research, Dubna, USSR* **1972,** *1972*.

44. *National Nuclear Data Center, NuDat database*. https://www.nndc.bnl.gov/nudat/ (accessed 01.10.2024).

45. Even, J.; Ballof, J.; Brüchle, W.; Buda, R. A.; Düllmann, Ch. E.; Eberhardt, K.; Gorshkov, A.; Gromm, E.; Hild, D.; Jäger, E.; Khuyagbaatar, J.; Kratz, J. V.; Krier, J.; Liebe, D.; Mendel, M.; Nayak, D.; Opel, K.; Omtvedt, J. P.; Reichert, P.; Runke, J.; Sabelnikov, A.; Samadani, F.; Schädel, M.; Schausten, B.; Scheid, N.; Schimpf, E.; Semchenkov, A.; Thörle-Pospiech, P.; Toyoshima, A.; Türler, A.; Vicente Vilas, V.; Wiehl, N.; Wunderlich, T.; Yakushev, A. The Recoil Transfer Chamber - An Interface to Connect the Physical Preseparator TASCA with Chemistry and Counting Setups. *Nucl. Instrum. Methods Phys. Res. A* **2011**, *638* (1), pp. 157–164. DOI: 10.1016/j.nima.2011.02.053.

46. Lens, L.; Yakushev, A.; Düllmann, Ch. E.; Asai, M.; Ballof, J.; Block, M.; David, H. M.; Despotopulos, J.; Di Nitto, A.; Eberhardt, K.; Even, J.; Götz, M.; Götz, S.; Haba, H.; Harkness-Brennan, L.; Heßberger, F. P.; Herzberg, R. D.; Hoffmann, J.; Hübner, A.; Jäger, E.; Judson, D.; Khuyagbaatar, J.; Kindler, B.; Komori, Y.; Konki, J.; Kratz, J. V.; Krier, J.; Kurz, N.; Laatiaoui, M.; Lahiri, S.; Lommel, B.; Maiti, M.; Mistry, A. K.; Mokry, C.; Moody, K.; Nagame, Y.; Omtvedt, J. P.; Papadakis, P.; Pershina, V.; Runke, J.; Schädel, M.; Scharrer, P.; Sato, T.; Shaughnessy, D.; Schausten, B.; Thörle-Pospiech, P.; Trautmann, N.; Tsukada, K.; Uusitalo, J.; Ward, A.; Wegrzecki, M.; Wiehl, N.; Yakusheva, V. Online Chemical Adsorption Studies of Hg, Tl, and Pb on $SiO_2$ and Au Surfaces in Preparation for Chemical Investigations on Cn, Nh, and Fl at TASCA. *Radiochim. Acta* **2018**, *106* (12), pp. 949–962. DOI: 10.1515/ract-2017-2914.

47. Götz, S.; Raeder, S.; Block, M.; Düllmann, Ch. E.; Folden, C. M.; Glennon, K. J.; Götz, M.; Hübner, A.; Jäger, E.; Kaleja, O.; Khuyagbaatar, J.; Kindler, B.; Krier, J.; Lens, L.; Lommel, B.; Mistry, A. K.; Mokry, C.; Runke, J.; Såmark-Roth, A.; Tereshatov, E. E.; Thörle-Pospiech, P.; Volia, M. F.; Yakushev, A.; Yakusheva, V. Rapid Extraction of Short-Lived Isotopes from a Buffer Gas Cell for Use in Gas-Phase Chemistry Experiments, Part II: On-line Studies with Short-Lived Accelerator-Produced Radionuclides. *Nucl. Instrum. Methods Phys. Res., B* **2021**, *507*, pp. 27–35. DOI: 10.1016/j.nimb.2021.09.004.

48. Węgrzecki, M.; Bar, J.; Budzyński, T.; Cież, M.; Grabiec, P.; Kozłowski, R.; Kulawik, J.; Panas, A.; Sarnecki, J.; Słysz, W.; Szmigiel, D.; Węgrzecka, I.; Wielunski, M.; Witek, K.; Yakushev, A.; Zaborowski, M. Design and properties of silicon charged-particle detectors developed at the Institute of Electron Technology (ITE). In *Electron Technology Conference 2013*; Szczepanski, P., Kisiel, R., Romaniuk, R. S., Eds.; SPIE, 2013, p 890212. DOI: 10.1117/12.2031041.

49. Khuyagbaatar, J.; Brand, H.; Düllmann, Ch. E.; Heßberger, F. P.; Jäger, E.; Kindler, B.; Krier, J.; Kurz, N.; Lommel, B.; Nechiporenko, Y.; Novikov, Y. N.; Schausten, B.; Yakushev, A. Search for fission from a long-lived isomer in $^{250}$No and evidence of a second isomer. *Phys. Rev. C* **2022**, *106* (2). DOI: 10.1103/PhysRevC.106.024309.

50. Hoffmann, J.; Kurz, N.; Loechner, S.; Minami, S.; Ott, W.; Rusanov, I.; Voltz, S.; Wieczorek, P. *New TASCA Data Acquisition Hardware Development for the Search of Element 119 and 120*: GSI Scientific Report PHN-IS-EE-02, pp. 253.

51. Kurz, N.; Hoffmann, J.; Minami, S.; Ott, W. *The MBS Data Aquisition System for the Search of Element 120 at TASCA*: GSI Scientific Report PHN-IS-EE-03, 253.

52. Even, J.; Yakushev, A.; Düllmann, Ch. E.; Dvorak, J.; Eichler, R.; Gothe, O.; Hartmann, W.; Hild, D.; Jäger, E.; Khuyagbaatar, J.; Kindler, B.; Kratz, J. V.; Krier, J.; Lommel, B.; Niewisch, L.; Nitsche, H.; Pysmenetska, I.; Schädel, M.; Schausten, B.; Türler, A.; Wiehl, N.; Wittwer, D. In-situ Formation, Thermal Decomposition, and Adsorption Studies of Transition Metal Carbonyl Complexes with Short-lived Radioisotopes. *Radiochim. Acta* **2014**, *102* (12), pp. 1093–1110. DOI: 10.1515/ract-2013-2198.

53. Eichler, R.; Schädel, M. Adsorption of Radon on Metal Surfaces: A Model Study for Chemical Investigations of Elements 112 and 114. *J. Phys. Chem. B* **2002**, *106* (21), pp. 5413–5420. DOI: 10.1021/jp015553q.

54. Soverna, S.; Dressler, R.; Düllmann, Ch. E.; Eichler, B.; Eichler, R.; Gäggeler, H. W.; Haenssler, F.; Niklaus, J.-P.; Piguet, D.; Qin, Z.; Türler, A.; Yakushev, A. B. Thermochromatographic Studies of Mercury and Radon on Transition Metal Surfaces. *Radiochim. Acta* **2005**, *93* (1), pp. 1–8. DOI: 10.1524/ract.93.1.1.58298.

55. Zvára, I. Simulation of Thermochromatographic Processes by the Monte Carlo Method. *Radiochim. Acta* **1985**, *38* (2), pp. 95–102. DOI: 10.1524/ract.1985.38.2.95.

56. Ziegler, J. F.; Biersack, J.; Ziegler, M. D. *SRIM - The Stopping and Range of Ions in Matter;* SRIM: Chester, Maryland, 2015.

57. Dietzel, D.; Yakushev, A.; Düllmann, Ch. E. An extended Monte Carlo simulation code for modeling gas chromatography experiments with superheavy elements and their homologs. *J. Radioanal. Nucl. Chem.* **2024** (333), pp. 3487–3496. DOI: 10.1007/s10967-023-09290-9.

58. Gilliland, E. R. Diffusion Coefficients in Gaseous Systems. *Ind. Eng. Chem.* **1934,** *26* (6), pp. 681–685. DOI: 10.1021/ie50294a020.